# A Hydrodynamic Model for Asymmetric Explosions of Rapidly Rotating Collapsing Supernovae with a Toroidal Atmosphere

V. S. Imshennik and K. V. Manukovskii*

*Institute for Theoretical and Experimental Physics,
ul. Bol'shaya Cheremushkinskaya 25, Moscow, 117259 Russia*

**Abstract**—We numerically solved the two-dimensional axisymmetric hydrodynamic problem of the explosion of a low-mass neutron star in a circular orbit. In the initial conditions, we assumed a nonuniform density distribution in the space surrounding the collapsed iron core in the form of a stationary toroidal atmosphere that was previously predicted analytically and computed numerically. The configuration of the exploded neutron star itself was modeled by a torus with a circular cross section whose central line almost coincided with its circular orbit. Using an equation of state for the stellar matter and the toroidal atmosphere in which the nuclear statistical equilibrium conditions were satisfied, we performed a series of numerical calculations that showed the propagation of a strong divergent shock wave with a total energy of $\sim 0.2 \times 10^{51}$ erg at initial explosion energy release of $\sim 1.0 \times 10^{51}$ erg. In our calculations, we rigorously took into account the gravitational interaction, including the attraction from a higher-mass $(1.9 M_\odot)$ neutron star located at the coordinate origin, in accordance with the rotational explosion mechanism for collapsing supernovae. We compared in detail our results with previous similar results of asymmetric supernova explosion simulations and concluded that we found a lower limit for the total explosion energy.

## INTRODUCTION AND FORMULATION OF THE HYDRODYNAMIC PROBLEM

Previously, we solved the two-dimensional axisymmetric problem of the formation of a toroidal atmosphere during the collapse of the rotating iron core and outer layers of a high-mass star by a numerical method identical to that used here (Imshennik *et al.* 2003). Our numerical calculations demonstrated the stability of the formed hydrostatically equilibrium toroidal atmosphere on characteristic hydrodynamic time scales—the calculations were performed up to times much longer than the hydrodynamic time of the problem (the final time of the main calculation is $t_f = 29.034$ s, while the characteristic hydrodynamic time is $t_{hd} = 0.517$ s). Our calculations on finer computational meshes (the number of mesh points in all directions was increased by factors of 1.5 and 2) revealed the possibility of sporadic fragmentation of this atmosphere even after a hydrostatic equilibrium was established. The fragmentation found took place once over the entire computational time and led only to a minor mass loss by the atmosphere. This fragmentation is most likely attributable to a restructuring of the toroidal atmosphere that results in its being roughly isentropic. This property of the atmosphere ($\nabla S \simeq 0$) and the distribution of specific angular momentum in it (which satisfies the condition $\partial j^2/\partial \tilde{r} > 0$, where $\tilde{r}$ is the cylindrical radius) derived previously (Imshennik *et al.* 2003) are necessary and sufficient conditions for the Viertoft–Lebowitz dynamical stability criterion to be fulfilled (Tassoul 1982). All of this allows the toroidal atmosphere obtained previously (Imshennik *et al.* 2003) to be deemed to be a long-lived structure with a total lifetime that is comparable, at least, to the evolution time of a neutron-star binary ($\gtrsim 1$ h) considered in the context of a rotational explosion mechanism for collapsing supernovae (Imshennik and Popov 1994; Imshennik and Ryazhskaya 2004).

Recall that the rotational mechanism implies a crucial role of the rotation effects in explaining the explosions of collapsing supernovae (Imshennik 1992). As was roughly estimated by Aksenov *et al.* (1997), for high-mass ($M > 10 M_\odot$) main-sequence stars, the equatorial rotational velocities of their iron cores at the final stages of their existence could be close in magnitude to the parabolic velocity. In the presence of such fast rotation, the iron core collapses to form a rapidly rotating protoneutron star that is generally unstable against the quadrupole dynamical rotational mode $m = 2$ (Aksenov *et al.* 1995) or, in other words, against its fragmentation. The neutron-star binary formed through such fragmentation (in the simplest case) evolves due to the losses of energy and angular momentum via the emission of gravitational waves, and the orbit of the binary becomes circular by the time the low-mass companion fills its Roche lobe almost independently of the initial eccentricity of the binary (Imshennik and Popov 1994). The filling of a Roche lobe with known characteristics (Paczynski 1971) leads to intense mass transfer from the less massive component to the more massive component of a neutron-star binary that ends with explosive destruction of the light component when it reaches the minimum possible neutron-star mass, $\simeq 0.1 M_\odot$ (Blinnikov *et al.* 1984, 1990; Colpi *et al.* 1989, 1991, 1993). It should be emphasized that the process of mass transfer in a neutron-star binary, which was considered in detail by Imshennik and Popov (1998), was simplified by using the so-called conservative approximation in which the possible formation of an accretion disk around the high-mass component was disregarded (Imshennik and Popov 2002). The existence of such an accretion disk during the mass transfer between the components of the binary under consideration has recently been analyzed by Colpi and Wasserman (2004). Based on the standard theory of such disks in close binaries (see Lubow and Shu 1975; Bildsten and Cutler 1992), they modeled the evolution of a close neutron-star binary in the opposite limiting case where all of the matter from the low-mass component is transferred to the accretion disk rather than is attached to the high-mass component. In the absence of reliable data on the characteristic dissipative disk destruction times, the analysis by Colpi and Wasserman (2002) is certainly of great interest. It not only does not restrict the rotational explosion mechanism for collapsing supernovae, but also enhances its capabilities through the emergence of a new scenario with earlier fragmentation of the rotating collapsar.

---

*E-mail:manu78@inbox.ru

Taking into account the aforesaid, we formulate below the axisymmetric hydrodynamic problem of the explosion of the low-mass component in a neutron-star binary that moves in a circular orbit in the presence of a rotating, hydrostatically equilibrium toroidal atmosphere with previously determined parameters (Imshennik *et al.* 2003). Below, we arbitrarily call the higher-mass component of the binary a pulsar (p) by also taking into account the possibility that this component turns into a black hole, which is not fundamentally important from the viewpoint of the problem considered here. The parameters of circular orbits in a binary are defined by Kepler's simple formulas. Thus, the orbital velocity of the low-mass neutron star (ns) is (Aksenov *et al.* 1997)

$$V_{\rm ns} = \frac{m_{\rm p}}{M_{\rm t}}(GM_{\rm t})^{1/2}a^{-1/2}, \quad (1)$$

where $M_{\rm t} = m_{\rm p} + m_{\rm ns}$ is the total mass of the binary, $m_{\rm p}$ is the mass of the pulsar, $m_{\rm ns}$ is the mass of the low-mass neutron star, and $a$ is the separation between the components. It is convenient to express the basic parameters of the problem in terms of the pulsar's velocity $V_{\rm p}$, which should be considered an observable parameter (Aksenov *et al.* 1997). The separation between the components is then

$$a = \left(\frac{m_{\rm ns}}{M_{\rm t}}\right)^2 \frac{GM_{\rm t}}{V_{\rm p}^2}. \quad (2)$$

The velocity and orbital radius of the low-mass neutron star can be determined from (1) and (2) and the obvious equality between the orbital periods of the two binary components

$$\frac{V_{\rm ns}}{V_{\rm p}} = \frac{a_{\rm ns}}{a_{\rm p}} = \frac{m_{\rm p}}{m_{\rm ns}}. \quad (3)$$

We take the results of the main numerical calculation from our previous paper (Imshennik *et al.* 2003) as the model of a toroidal atmosphere. This calculation was performed in a spherical layer whose inner boundary, as can be easily verified, is located along the radius outside the region of space occupied by the neutron-star binary. The toroidal atmosphere computed previously (Imshennik *et al.* 2003) definitely lies outside this region. We also took the total mass of the binary $M_{\rm t}$ from this calculation; the sum of the masses of the protoneutron-star embryo ($\sim 1 M_\odot$) and the matter accreted through the inner computational boundary onto this embryo over the formation time of the toroidal atmosphere should be taken as this mass. Meeting this requirement ensures an approximate satisfaction of the hydrostatic equilibrium conditions for the toroidal atmosphere. To completely determine the parameters of the binary, it remains only to specify the pulsar's velocity and the relationship between the masses of the pulsar and its lighter companion.

A similar problem of the explosion of a low-mass neutron star was studied by Aksenov *et al.* (1997) and Imshennik and Zabrodina (1999). Nevertheless, these studies had a number of fundamental differences from the problem considered here. First, there is a difference in the formulations of the hydrodynamic problem, although the numerical simulations are two-dimensional and axisymmetric in both cases. In the papers mentioned above, the axis of symmetry of the problem coincided with the velocity vector of the rotating low-mass neutron star in a circular orbit at the time of its explosion. The gravitational interaction was completely ignored. Clearly, the latter assumptions are asymptotically valid at distances $r \gg a$. Aksenov *et al.* (1997) interpreted this formulation of the problem as the artificial turn of the velocity direction for the exploding star through $\pi/2$, and the estimates obtained confirmed that the gravitational interaction effects were negligible. However, these effects could be rigorously taken into account only in the three-dimensional formulation of the problem (see the next section), although, as was noted above, the problem asymptotically becomes two-dimensional at $r \gg a$. In this paper, as in the preceding paper on the formation of a toroidal atmosphere, we also use the two-dimensional axisymmetric approximation, but the axis of symmetry coincides with the axis of orbital rotation of the neutron-star binary (rather than is perpendicular to it!). In this formulation of the problem, the only way of reducing the fundamentally three-dimensional problem of the explosion of a low-mass neutron star to a two-dimensional problem is to "spread" the rotating neutron star over its orbit or, in other words, to represent it as an exploding torus with a circular cross section. In fact, as Aksenov's preliminary calculations show[1], the light component of the binary traverses less than a quarter of the orbital circumference since the onset of destruction before leaving it. It may be asserted that replacing the low-mass neutron star with an exploding torus is a forced, but quite relevant approximation, especially since a significant difference of this work is the direct allowance for the pulsar's gravitational influence, along with the allowance for the self-consistent gravitational field produced by the distributed matter throughout the computational region by directly solving the Poisson equation (see below). As was noted above, Aksenov *et al.* (1997) and Imshennik and Zabrodina (1999) disregarded the gravitational interaction in their numerical solution.

Another significant difference of the hydrodynamic problem in question from the previous problems is the use of spherical coordinates, whereas the above authors performed their numerical calculations in cylindrical coordinates. In addition, Aksenov *et al.* (1997) used computational meshes of two types, adaptive (LM) and fixed (PPM), to mutually check the accuracy, while Imshennik and Zabrodina (1999) used only adaptive (LM) meshes, which are recognized to be more adequate. Finally, matter with the simplest (uniform) density distribution was located on the path of the divergent shock wave in the above papers. In this paper, a toroidal atmosphere with a nonuniform density distribution was naturally located on the path of the shock wave.

---

[1]We are grateful to A.G. Aksenov for this private communication.

## INITIAL CONDITIONS. THE HYDRODYNAMIC MODEL

At the end of the numerical solution ($t_f = 29.034$ s), the toroidal atmosphere obtained previously (Imshennik *et al.* 2003) (the main calculation [1]) had the following parameters (see Table 3 in the above paper): the maximum density is $\rho_{\max} = 0.396 \times 10^7$ g cm$^{-3}$, the position of the density maximum is a point on the equator (in the cylindrical ($\tilde{r}, z$) coordinate plane) with a radius of $\tilde{r}_{\max} = 0.955 \times 10^8$ cm, and the total mass of the atmosphere is $M_{\text{atm}} = 0.117 M_\odot$. The inner computational boundary in calculation [1] was located at a spherical radius of $r^*_{\min} = 0.876 \times 10^8$ cm, which, as was shown previously (Imshennik *et al.* 2003), may be considered as the inner boundary of the toroidal atmosphere. The mass of the matter located below the radius $r^*_{\min}$ (outside the computational region) at the final time $t_f$ was $M_{\text{in}} = M_t = 1.931 M_\odot$. Note that $M_t$ agrees well with the typical iron core masses of high-mass stars, $M_{\text{Fe}}$ ($1.2 M_\odot < M_{\text{Fe}} < 2 M_\odot$), and surprisingly closely matches the value of $M_t = 1.9 M_\odot$ used by Aksenov *et al.* (1997).

For the pulsar's velocity $V_p$, we take a reasonable value (Lyne and Lorimer 1994):

$$V_p = 1000 \text{ km s}^{-1}. \quad (4)$$

We take the mass of the exploding torus (the low-mass neutron star at the time of its destruction) from Aksenov *et al.* (1997); i.e., we assume that $m_{\text{ns}} = 0.1 M_\odot$. According to (2), the separation between the components of the binary is then $a = 6.98 \times 10^7$ cm $\approx 700$ km, which is below identified with the orbital radius of the low-mass neutron star: $a_{\text{ns}} = a$. The radius of the torus $r_t$ can be determined by assuming that the volumes of the neutron star (a sphere of radius $r_0$) and the torus with a circular cross section (in the approximation $r_t \ll a$) are equal:

$$V_{\text{torus}} \simeq 2\pi a \pi r_t^2 = \frac{4}{3}\pi r_0^3. \quad (5)$$

As $r_0$, Aksenov *et al.* (1997) took $0.1 R_{\text{Fe}}$, where $R_{\text{Fe}} = 4.38 \times 10^8$ cm is the initial radius of the iron core. The radius of the circular torus is then

$$r_t = \left(\frac{2 r_0^3}{3\pi a}\right)^{1/2} = 1.60 \times 10^7 \text{ cm} = 160 \text{ km}. \quad (6)$$

It thus follows that the low-mass neutron star considered in the form of a circular torus at the onset of its explosion is contained within a sphere with a radius of $\sim 860$ cm, which is slightly smaller than the radius of the inner boundary of the toroidal atmosphere identified above with the radius of the computational region $r^*_{\min}$ (Imshennik *et al.* 2003). Apart from the representation of the low-mass neutron star as a torus, we assume that the higher-mass component of the binary (a pulsar) is at the coordinate origin, i.e., $a_p = 0$, which naturally agrees with the above equality $a_{ns} = a$. An additional justification for this assumption is the strong inequality $V_p/V_{ns} = a_p/a_{ns} = 1/18 \approx 0.0556 \ll 1$, as implied by (3). The satisfaction of the hydrostatic equilibrium condition for a circular torus in the gravitational field of a stationary pulsar is also appropriate in the initial conditions for the hydrodynamic problem under consideration. This requires determining the velocity of the low-mass neutron star using the formula $V_{ns} = (Gm_p/a)^{1/2}$, which yields $18.5 \times 10^3$ km s$^{-1}$ instead of $18 \times 10^3$ km s$^{-1}$ obtained from (3) and (4).

Here, it is pertinent to independently estimate the gravitational interaction between the matter of the exploded low-mass neutron star and the higher-mass component. This can be done by following the three-dimensional model in the dust approximation by Colpi and Wasserman (2002) in the limit $m_{ns} \ll m_p$ that holds in the case under consideration. According to these authors, the pulsar's kick velocity is

$$V_{kick} = \eta \frac{m_{ns}}{m_p} V_{expl},$$

where $V_{expl} = 18.5 \times 10^3$ km s$^{-1}$ is the orbital velocity of the exploded star with respect to the center of inertia, $\eta = \eta(w_0')$ is the gravitational deceleration coefficient, and $w_0' = w_0/(Gm_p/a)^{1/2}$ with $w_0 = (2E_0/m_0)^{1/2} = 3.01 \times 10^9$ cm s$^{-1}$ at $E_0 = 4.7$ MeV/nucleon ($m_0 = 1.66 \times 10^{-24}$ g). At $m_p = 1.8 M_\odot$ and $a = 7 \times 10^7$ cm, $w_0' = 1.63$ and, accordingly, $\eta = 0.73$ (see Fig. 1 from Colpi and Wasserman 2002). It thus immediately follows that $V_{kick} = 750$ km s$^{-1}$ instead of $V_{kick} = 1000$ km s$^{-1}$ from (4). Thus, the initial kick velocity of the pulsar decreases, although only slightly, and still agrees with the observational data on the high velocities of pulsars (Lyne and Lorimer 1994).

Let us consider the choice of initial conditions in the region of energy release in more detail. This question was discussed in detail by Imshennik and Zabrodina (1999) and Zabrodina and Imshennik (2000), who made important refinements. In their hydrodynamic calculation of the explosive destruction of a self-gravitating neutron star with a critical mass, Blinnikov et al. (1990) obtained the internal energy of the explosion products, $E_0 = 4.70$ MeV/nucleon $= 4.5 \times 10^{18}$ erg g$^{-1}$. Below, we use this value as the basis for determining the specific energy release[2]. According to the equation of state used here (see the next section), part of the internal energy of the matter is contained in the rest energy of the nuclides (if the iron mass fraction $X_{Fe}$ is less than unity). Below, the specific internal energy minus this part is denoted by $e_0$. In addition, $E_0$ can decrease appreciably due to neutrino radiation and allowance for the final times of $\beta$-processes. Therefore, it is appropriate to use $\xi E_0$, where $\xi \leq 1$ is the explosion attenuation coefficient, in place of $E_0$. Formally, this coefficient may be set larger than unity ($\xi > 1$), bearing in mind the possibility of errors in the quantity $E_0$ itself in the cited paper. Such cases will also be presented below. The coefficient $\alpha$ that specifies the initial value of the specific internal energy of the matter in the region of energy release introduced above, $e_0 = \alpha E_0$, can then be estimated using the formula (Zabrodina and Imshennik 2000)

$$\alpha = \xi - 2.012(1 - X_{Fe}) + 1.654 X_{He} + 0.2771 X_p, \tag{7}$$

where $X_{Fe}$, $X_{He}$, and $X_p$ are the mass fractions of $^{56}_{26}$Fe, $^{4}_{2}$He, and protons, respectively, which are known functions of the thermodynamic quantities $e_0$ and $\rho_0$. For $X_{Fe} = 1$ ($X_{He} = X_p = 0$), we obtain $\alpha = \xi$ from (7), as would be expected. Thus, to find the initial thermodynamic state of the low-mass neutron star, we must determine the coefficient $\alpha = \alpha(\xi, e_0, \rho_0)$ using (7) by numerically solving the equation

$$e_0 = \alpha(\xi, e_0, \rho_0) E_0. \tag{8}$$

Under the assumption of a uniform initial density distribution over the entire circular torus of mass $m_{ns} = 0.1 M_\odot$ with the parameters $a$ and $r_t$ specified above, i.e., for the mean density $\rho_0 = 5.66 \times 10^8$ g cm$^{-3}$ and the coefficient $\xi = 1$, Eq. (8) yields the initial internal energy of the neutron-star matter $e_0 = 3.14 \times 10^{18}$ erg g$^{-1}$ (see Table 1), which determines all of the remaining initial-state parameters for the region of energy release (the first column). This result was previously obtained by Zabrodina and Imshennik (2000) (see the table in the cited paper) and, naturally, closely matches the above value of $e_0$ for the equation of state used in this paper. For a uniform distribution of the density and other thermodynamic parameters, the total internal energy of the neutron star is $\varepsilon_0 = e_0 m_{ns} = \alpha m_{ns} E_0$. For a nonuniform distribution, it was obtained by integration over the entire volume

---

[2] It should be borne in mind that the equation of state in the cited paper was a simple interpolation between the equation of state for cold catalyzed matter $P_0(\rho)$ and $E_0(\rho)$ (Baym et al. 1971) and the equation of state for an ideal gas as a temperature additive in place of the equation of state for nonideal nuclear matter with a nonzero temperature (Lattimer and Swesty 1989).

of the torus, which pertains to the second and third columns in Table 1.

For the subsequent numerical solution, we had to fill the space of the iron-core cavity around the exploded neutron star (down to the radius $r^*_{\min}$) with low-density ($10^6$ g cm$^{-3}$) matter. To reduce the initial jump in density that arises in this case at the contact discontinuity between th explosion products and the matter of the iron-core cavity, the initial distribution of thermodynamic quantities was smoothed out near the boundary of the circular torus within which the density distribution became nonuniform. To keep the total mass of the low-mass neutron star constant ($m_{\mathrm{ns}} = 0.1 M_\odot$), the central density of the circular torus was artificially increased by a factor of 1.5 ($\rho'_0 = 8.36 \times 10^8$ g cm$^{-3}$). Naturally, this entailed changes in other initial parameters in the region of energy release. This set of, strictly speaking, central parameters (with the same $e_0$) was chosen as the initial state of the exploding neutron star in the main calculation (the second column in Table 1). A comparison of the first two columns in Table 1 shows that the mass fractions of the nuclides $X_{\mathrm{Fe}}$, $X_{\mathrm{He}}$, $X_{\mathrm{p}}$, and $X_{\mathrm{n}}$ were almost identical, while the parameters $\alpha$ and $\xi$, which characterize the degree of energy release, proved to be equal to their previous values, to within small corrections. The initial pressure $P_0$ and temperature $T_0$ are slightly higher in the second column. Nevertheless, the total internal energy of the neutron star in the case under consideration proved to be even slightly lower ($\varepsilon_0 = 0.53 \times 10^{51}$ erg). As a result, these changes in the initial conditions of the problem seem negligible.

The third column of Table 1 gives a set of initial parameters for the formal case $\xi = 1.84$ (the case of highly overestimated initial energy release with $\xi E_0 = 8.28 \times 10^{18}$ erg g$^{-1}$), which was used in our auxiliary calculation (see the section "Discussion of Numerical Results"). In this case, according to the chosen equation of state, the matter of the low-mass neutron star initially consists of predominantly helium and a small fraction of free neutrons, while the iron mass fraction $X_{\mathrm{Fe}}$ is equal to zero. In essence, this case ignores the expenditure of energy on overcoming the self-gravity of the exploding neutron star, but is still consistent with the total energy release during the recombination of neutron matter into iron, $E_0 \approx 9.2$ MeV/nucleon $= 8.8 \times 10^{18}$ erg g$^{-1}$.

## THE METHOD OF NUMERICAL SOLUTION

When the explosion of a low-mass neutron star is modeled, the system of ideal hydrodynamic equations in the axisymmetric case ($\partial/\partial\varphi, g_\varphi = 0$) in spherical coordinates ($r, \theta, \varphi$) that was described in detail previously (Imshennik et al. 2002) is solved numerically.

**Table 1.** Initial-state parameters for the region of energy release at various central densities $\rho_0$ and specific internal energies $e_0$

| | | | |
|---|---|---|---|
| $\zeta$ | 1.00 | 1.01 | 1.84 |
| $\alpha$ | 0.700 | 0.701 | 1.186 |
| $\varepsilon_0$, ($10^{51}$ erg) | 0.63 | 0.53 | 0.90 |
| $e_0$, ($10^{18}$ erg g$^{-1}$) | 3.14 | 3.14 | 5.31 |
| $\rho_0$, ($10^8$ g cm$^{-3}$) | 5.66 | 8.36 | 8.36 |
| $T_0$, ($10^9$ K) | 8.88 | 9.63 | 11.95 |
| $P_0$, ($10^{26}$ erg cm$^{-3}$) | 3.73 | 6.07 | 8.79 |
| $X_{\mathrm{Fe}}$ | 0.385 | 0.370 | 0.000 |
| $X_{\mathrm{He}}$ | 0.566 | 0.579 | 0.814 |
| $X_{\mathrm{n}}$ | $4.65 \times 10^{-2}$ | $4.79 \times 10^{-2}$ | 0.123 |
| $X_{\mathrm{p}}$ | $2.58 \times 10^{-3}$ | $2.95 \times 10^{-3}$ | $5.74 \times 10^{-2}$ |

We numerically solved the system of hydrodynamic equations using an algorithm that is based on the PPM method (Colella and Woodward 1984) and that is a modification of Godunov's method (Godunov et al. 1976). The method has been repeatedly described previously (see, e.g., Imshennik et al. 2002).

The gravity $\mathbf{g} = \mathbf{g}_{\mathrm{p}} + \mathbf{g}_{\mathrm{env}}$ that acts on the matter in the problem under consideration is the sum of two parts: the first is attributable to the gravitational field of the pulsar placed exactly at the coordinate origin:

$$\mathbf{g}_{\mathrm{p}} = \left(-\frac{Gm_{\mathrm{p}}}{r^2}, 0, 0\right), \qquad (9)$$

while the second is attributable to the gravitational field of the matter of the computational region, whose gravity is defined by the standard equation $\mathbf{g}_{\mathrm{env}} = -\nabla\Phi$, and the potential satisfies the Poisson equation

$$\Delta\Phi = 4\pi G\rho. \qquad (10)$$

An efficient algorithm designed for use on stationary meshes in spherical coordinates is used to solve the Poisson equation (10) and to determine the gravity $\mathbf{g}_{\mathrm{env}}$. This algorithm is based on the expansion of the integral representation of the gravitational potential $\Phi$ in terms of associated Legendre polynomials (Aksenov 1999). In our calculations, the potential was expanded in terms of the first twenty associated Legendre polynomials. The necessary boundary condition for the gravitational potential, $\Phi \to -(GM)/r$ for $r \to \infty$, is automatically satisfied in this algorithm of solving the Poisson equation.

Here, the matter over the entire temperature range under consideration is assumed to be a mixture of an ideal Boltzmann gas of free nucleons n, p and nuclides $^{4}_{2}$He, $^{56}_{26}$Fe as well as an ideal Fermi–Dirac electron–positron gas together with equilibrium blackbody radiation. The equation of state is subject to the nuclear statistical equilibrium conditions with a fixed ratio of the mass fractions of the neutrons and protons, including those bound in the helium and iron nuclides, $\theta = 30/26$. The specific internal energy of the matter is defined by the formula (see Eq. (9) from Imshennik and Zabrodina (1999))

$$e = e_- + e_+ + e_r + e_{id} \quad (11)$$
$$+ \left[ \frac{Q_{Fe} + 26\Delta Q_n}{56 m_u}(1 - X_{Fe}) \right.$$
$$\left. - \frac{Q_{He} + 2\Delta Q_n}{4 m_u} X_{He} - \frac{\Delta Q_n}{m_u} X_p \right],$$

where $e_-$ and $e_+$ are the contributions of the electrons and positions to the internal energy of the matter, respectively; $e_r$ is the contribution of the blackbody radiation; and $e_{id}$ is the contribution of the ideal Boltzmann gas of nuclides. The expression in the square brackets is the part of the internal energy of the matter that is contained in the rest energy of the nuclides and that has a direct bearing on the choice of an initial state for the exploding neutron star (see above). In the solution, the equation of state is tabulated, which increases appreciably the speed of the numerical method. In addition, the dependence of the matter pressure on density $\rho$ and specific internal energy $e$ is locally simulated by a binomial approximation (see, e.g., Imshennik et al. 2003).

The system of units from our previous paper (Imshennik et al. 2003) is used in the numerical solution. The numerical scales of the physical quantities in this system are

$$[r] = 10^8 \text{ cm}, \quad [M] = 10^{32} \text{ g}, \quad (12)$$
$$[V_r] = [V_u] = [V_\varphi] = 2.583 \times 10^8 \text{ cm s}^{-1},$$
$$[c] = 7.958 \times 10^6 \text{ g}^{-3}, \quad [t] = 3.871 \times 10^{-1} \text{ s},$$
$$[P] = 5.310 \times 10^{23} \text{ erg cm}^{-3},$$
$$[E] = 6.674 \times 10^{16} \text{ erg}, \quad [T] = 2.894 \times 10^9 \text{ K}.$$

The region of solution of the problem or the computational region is in the shape of a spherical envelope with $r_{min} \leq r \leq r_{max}$, $r_{min} = 5 \times 10^7$ cm, $r_{max} = 1.000168 \times 10^9$ cm. The inner boundary at the radius $r = r_{min}$ is assumed to be transparent, which is roughly achieved by setting the gradients of all physical quantities ($\mathbf{V}, \rho$, and $e$) equal to zero in the radial direction. At the radius $r = r_{max}$, the boundary condition simulates a vacuum outside the computational region: (nearly zero) background values are assigned to the thermodynamic quantities ($\rho, e$, and $P$). The boundary conditions are sufficient for the difference scheme used, and their influence on the solution is assumed to be negligible.

In view of the equatorial symmetry along with the axial symmetry, it will suffice to find the solution only in one quadrant. Therefore, the angle $\theta$ for the computational region varies over the range 0 to $\pi/2$. At the boundary $\theta = \pi/2$, the velocity component $V_\theta$ is set equal to zero (and again the derivatives of all thermodynamic quantities become equal to zero). Thus, the choice of boundary conditions does not differ in any way from their choice in our previous papers (Imshennik et al. 2002, 2003).

## DISCUSSION OF NUMERICAL RESULTS

In all probability, the following three calculations represent our main results most completely. Calculations [1] and [2] use the set of values from the second column of Table 1 as the initial state of the low-mass neutron star. Calculation [3] models an explosion with significantly overestimated energy release (the third column of Table 1). In all our calculations, we used the same computational mesh. The outer boundary of the computational region was located at the dimensionless radius $r_{max} = 10.00168$, whose choice was discussed previously (Imshennik et al. (2003). There was arbitrariness in choosing the location of the inner computational boundary. The dimensionless radius in all our calculations was $r_{min} = 0.5$, so the low-mass neutron star modeled in the form of a torus with a circular cross section (see above) was completely within the computational region. The spherical layer from $r_{min}$ to $r^*_{min} = 0.876$ was broken down into 40 equal zones in the radial direction. The computational mesh in the region from $r^*_{min}$ to $r_{max}$ was taken from our previous main calculation (Imshennik et al. 2003) (100 zones in the radial direction), which made it unnecessary to recalculate the spatial distribution of thermodynamic quantities in an equilibrium toroidal atmosphere for the new problem. The total number of zones in the direction in which the polar angle changed was 30 in all our calculations. A distinctive feature of calculation [2] is the replacement of an equilibrium toroidal atmosphere by an atmosphere with uniform density and temperature distributions with the integrated parameters of the matter (its total mass and internal energy) kept constant.

The numerical solution in all our calculations was performed up to a time of $\sim 1.3$ s, i.e., longer than the calculations by Aksenov et al. (1997) and Imshennik and Zabrodina (1999). Figures 1–3 present the results of these calculations for a certain characteristic time, $t = 0.4$ s, whose choice was justified by the convenience of a comparison with our previous calculations (to be more precise, the main results in these papers are presented for a time of $t \approx 0.43$ s). Figures 1a–1c show the lines of constant logarithm of the dimensionless density for calculations [1–3], respectively. We see from these figures that the shape of the shock front generated by the explosion of a low-mass neutron star deviates from a sphere, particularly for calculations [1] and [3]. In contrast, this deviation is less pronounced for a homogeneous atmosphere (calculation [2]). For calculation [1], the velocities of the shock front characterized by a large velocity gradient in the equatorial plane and along the rotation axis differ by more than a factor of 1.5 and are equal to $\sim 1.2 \times 10^9$ and $\sim 2 \times 10^9$ cm s$^{-1}$, respectively. By the time under consideration, the leading part of the shock front (the upper part of Fig. 1a) reaches the outer spherical boundary of the computational region. In calculation [3], this part of the shock front is already outside the computational region by this time ($t = 0.4$ s). The current locations of the shock front on the equator at the time under consideration in calculations [2] and [3] almost coincide and correspond to a radius of $r_{sw} \approx 7.8$, while in calculation [1], $r_{sw} \approx 6.6$.

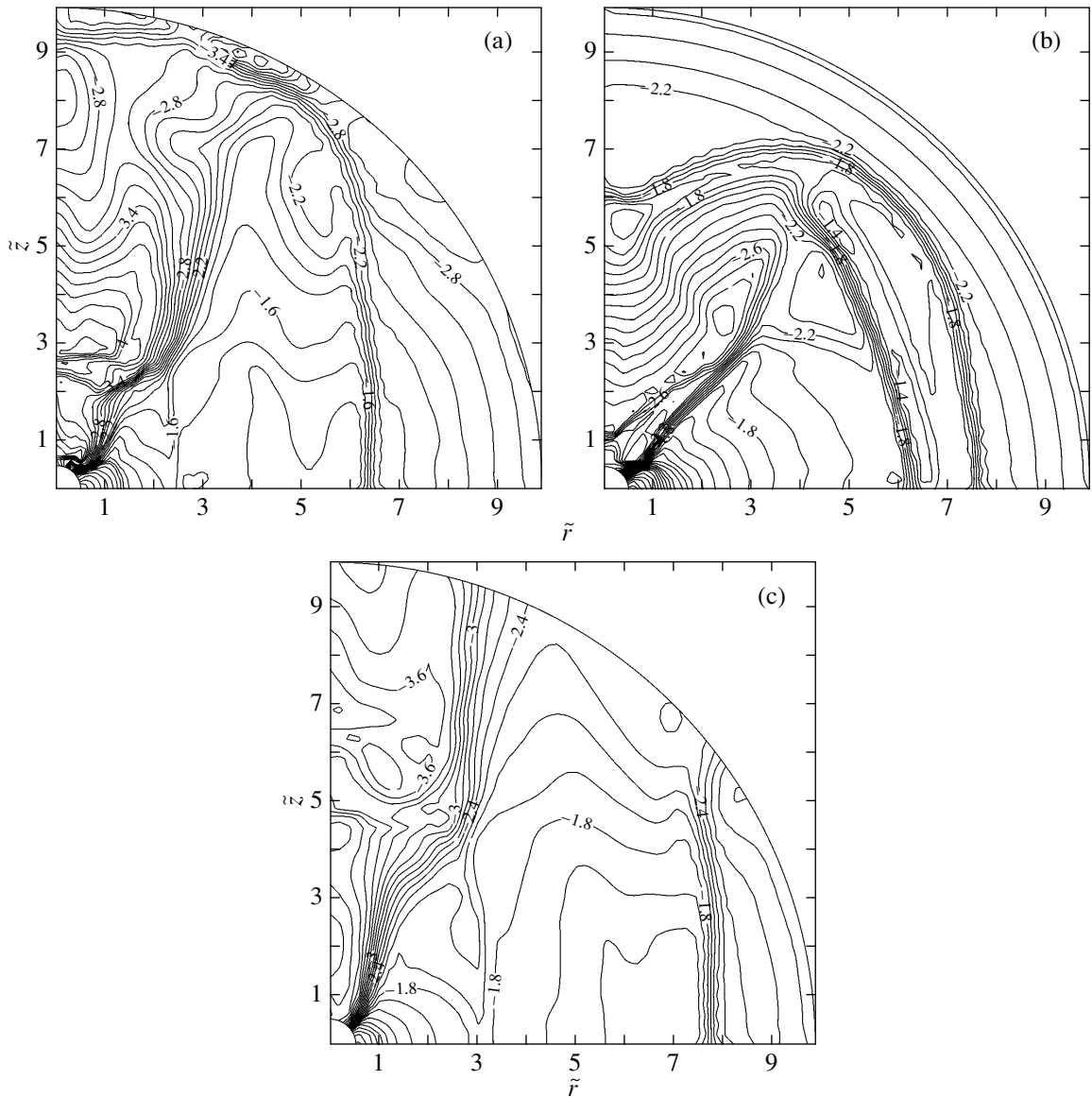

**Fig. 1.** Lines of constant logarithm of the density $\log \rho$ as a function of the cylindrical coordinates $\tilde{r}$ and $\tilde{z}$ for calculations [1] (a), [2] (b), and [3] (c) and for the time $t = 0.4$ s.

In the main calculation by Imshennik and Zabrodina (1999) (see Fig. 4 from their paper) with the corresponding initial internal energy $\varepsilon_0 = 0.675 \times 10^{51}$ erg (calculation II of Table 1 from their paper), the shock front reached a radius of $r_+ = z_{\max} \approx 12$ in the leading direction by this time, while the corresponding radius in the trailing direction is $r_- = |z_{\min}| \approx 3.0$. Thus, the mean radius of the divergent and almost spherical shock front was $r_{\rm sw} = (r_- + r_+)/2 \approx 7.5$. In their main calculation, Aksenov et al. (1997) (see Fig. 8 from their paper) obtained a slightly larger value of $r_{\rm sw} \approx 8.2$, but their equation of state differs significantly from that used here and by Imshennik and Zabrodina (1999), although the final values of the energy release ($0.9 \times 10^{51}$ erg) in these calculations were approximately equal between themselves and to those in our calculations [1] and [2]. It seems natural that the radii from the previous papers being compared quantitatively matched most closely the radius in calculation [2] with a uniform density distribution, and not in calculation [1] with a nonuniform density distribution of the toroidal atmosphere. In the latter case, however, there is good agreement with the model of a weak explosion in the paper by Zabrodina and Imshennik (2000), where the lower initial internal energy ($\varepsilon_0 = 0.38 \times 10^{51}$ erg) corresponded to the decrease in final energy release by exactly a factor of 2 ($\varepsilon_0 = 0.45 \times 10^{51}$ erg). Indeed, according to Fig. 2 from Zabrodina and Imshennik (2000), $r_+ \approx 12$, while $r_- \approx 2.2$ and, hence, $r_{\rm sw} \approx 7.1$.

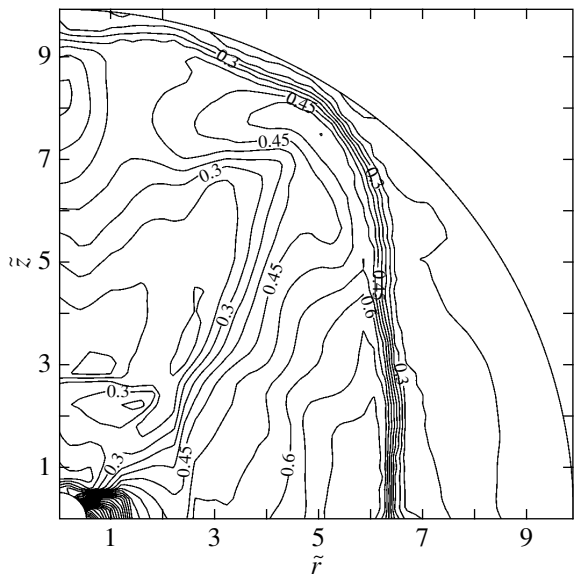

**Fig. 2.** Lines of constant temperature $T$ as a function of the cylindrical coordinates $\tilde{r}$ and $\tilde{z}$ for calculation [1] and for the time $t = 0.4$ s.

In all three calculations, the shock front reaches the radius $r_{\max}$ in the equatorial plane approximately at the same time, $\sim 0.6$–$0.7$ s, with a small delay of 0.1 s for calculation [1]. In contrast to calculations [1] and [3], in which an extended region with almost constant density and temperature distributions is formed behind the shock front, a spherical layer of dense hot matter is formed in calculation [2]. Figure 2 shows the temperature distribution for calculation [1] at the time $t = 0.4$ s, which, as can be seen from the figure, resembles the density distribution, with the only difference that the lines of constant temperature behind the shock front are directed predominantly along the rotation axis (the temperature changes mainly in the equatorial direction), while the lines of constant density are more likely parallel to the equatorial plane.

Figure 3 shows the profiles of the logarithm of the density, temperature, and radial velocity for calculation [1] as a function of the cylindrical radius $\tilde{r}$ for $\theta = \pi/2$ (in the equatorial plane) at consecutive times from $t = 0$ to $t = 0.7$ s. At the initial time, the densities and temperatures are hind inside the circular torus (Figs. 3a and 3b). As we pointed out above, almost constant densities and temperatures are established behind the shock front at later times ($t > 0.4$ s), while at $t < 0.4$ s, the distributions exhibit minima that roughly correspond to the contact boundary between the matter of the toroidal atmosphere and the explosion products of the low-mass neutron star. In addition, we see from Fig 3b that, starting from a time of $t \approx 0.1$ s, the temperature of the toroidal atmosphere does not exceed its critical value $T_{\rm cr}$ that is determined by the well-known theoretical temperature minimum ($3$–$5 \times 10^9$ K) for the approximation of nuclear statistical equilibrium to be applicable. Imshennik and Zabrodina (1999) assumed $T_{\rm cr}$ to be $4.17 \times 10^9$ K. For convenience, we will use a close dimensionless value of $T_{\rm cr} = 1.5$. It also follows from the computed data on the chemical composition that all of the matter in the computational region almost completely recombines into iron by the time $t = 0.1$ s in all our calculations, including calculation [3] of a strong explosion. This once again confirms the previous conclusion that the effect of iron dissociation into free nucleons is negligible (see, e.g., Zabrodin and Imshennik 2000).

The velocity of the matter behind the shock front decreases to $\sim 9 \times 10^8$ cm s$^{-1}$ by a time of $t \approx 0.3$ s and subsequently increases only slightly to $10^9$ cm s$^{-1}$ (see Fig. 3c). The velocity of the matter in calculation [3] behaves similarly, changing over a narrow range, $(1.3$–$1.5) \times 10^9$ cm s$^{-1}$. In contrast, its behavior in calculation [2] differs markedly: the velocity is high at the initial expansion phase, $\sim 1.8 \times 10^9$ cm s$^{-1}$, but subsequently, starting from a time of $t \approx 0.4$ s, it rapidly decreases by a factor of 2 to $10^9$ cm s$^{-1}$. In all our calculations, a moderate rarefaction wave is formed in the region adjacent to the inner computational boundary. The maximum of the absolute velocity of the matter in the rarefaction wave is about $10^9$ cm s$^{-1}$ and lies well to the right from the radius $\tilde{r}_{\max}$ for the initial maximum of the atmospheric density.

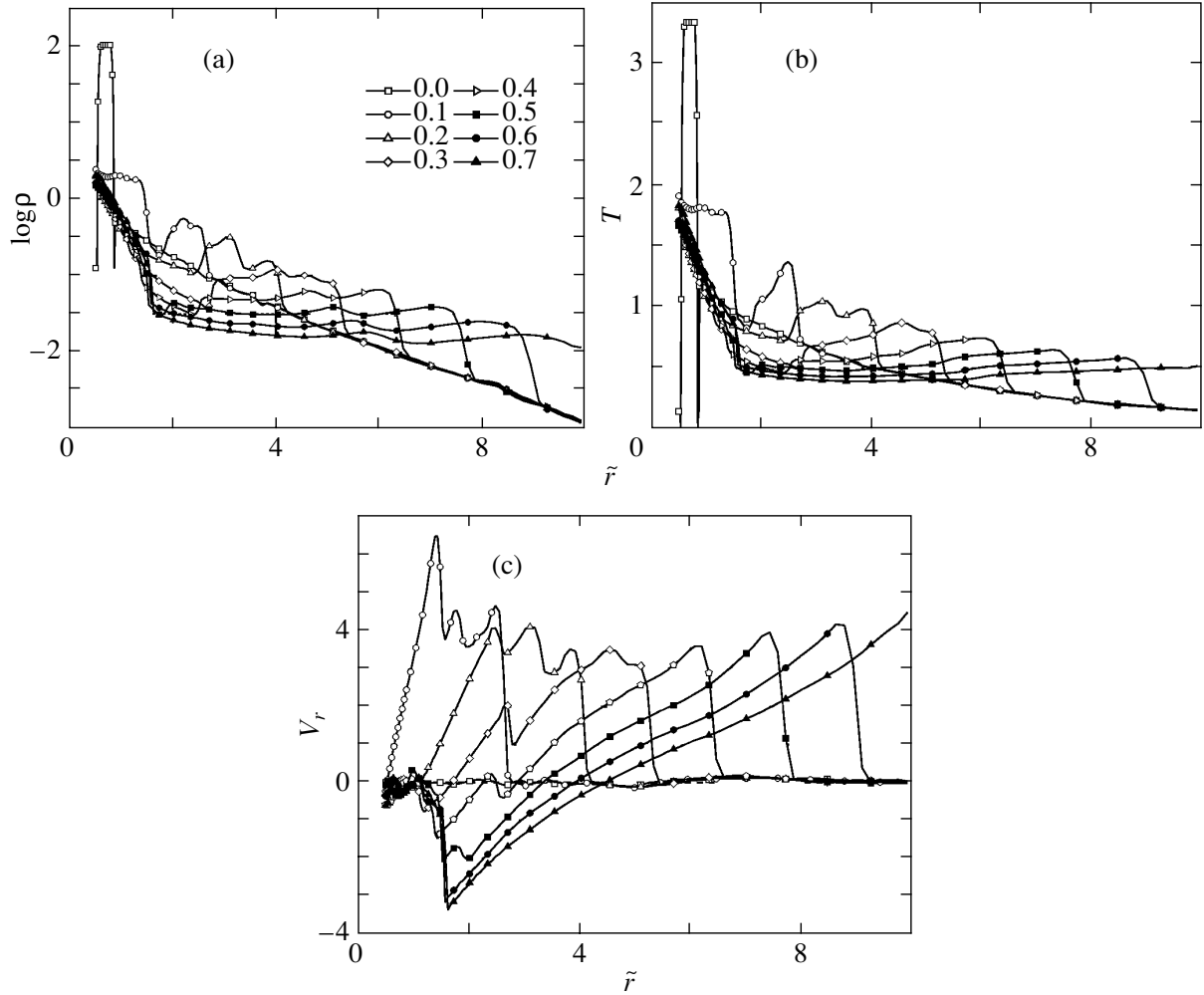

**Fig. 3.** Profiles of the logarithm of the density $\log \rho$ (a), temperature $T$ (b), and radial velocity $V_r$ (c) as a function of the cylindrical radius $\tilde{r}$ at $\theta = \pi/2$ for calculation [1] at sequential times from the beginning of our calculation $t = 0$ up to $t = 0.7$ s.

In Fig. 4, the integral of the total energy flux ($\varepsilon_{\rm tot}$) through the outer boundary of the computational region is plotted against time. We calculated $\varepsilon_{\rm tot}$ as the sum of the total internal and kinetic energies of the matter. The kinetic energy also includes the rotational kinetic energy; however, its contribution is negligible. As we see from Fig. 4, the curves reach constant values starting from a time of $t \simeq 1.0$ s. The asymptotic values of the total energy were found to be $2.3 \times 10^{50}$, $2.6 \times 10^{50}$, and $5.5 \times 10^{50}$ erg for calculations [1–3], respectively. Thus, the total energy $\varepsilon_{\rm tot}$ for calculation [1], which is most justifiable in terms of the final energy release, is appreciably lower than the characteristic total energy of supernova explosions ($\sim 10^{51}$ erg). At the same time, in general, the problem of a deficit in total energy did not arise in the hydrodynamic calculations of an asymmetric explosion (Zabrodina and Imshennik 2000). In this paper, the initial kinetic energy of the low-mass neutron star ($E_k \approx 3.5 \times 10^{50}$ erg) is entirely contained in the azimuthal velocity $V_\varphi$, which is specified in the initial data in such a way as to balance the exploding torus in the pulsar's gravitational field. In contrast, the orbital kinetic energy of the low-mass component of the binary in the hydrodynamic model of an asymmetric explosion (Aksenov et al. 1997; Imshennik and Zabrodina 1999) was essentially an appreciable contribution to the total energy release. In addition, the gravitational interaction with the pulsar, which hampered the development of an explosion to some (small) extent, was disregarded in the previous papers. Then, it should probably recognized that the explosion energy was overestimated, most likely only slightly, in that model. In contrast, the value obtained here, $\varepsilon_{\rm tot} = 2.3 \times 10^{50}$ erg, should be considered as a lower limit on the total energy of an asymmetric supernova explosion for given initial energy release (within the framework of the rotational mechanism).

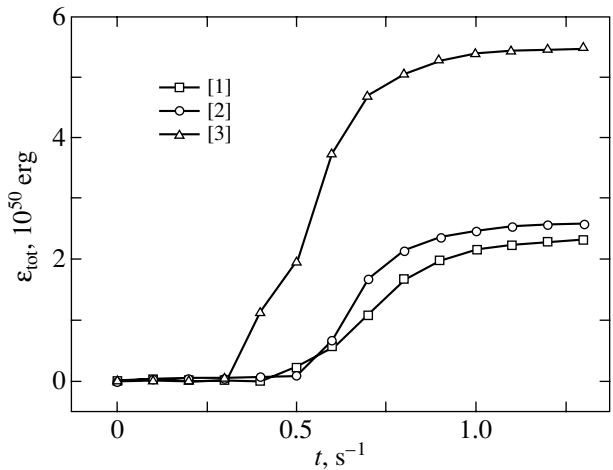

**Fig. 4.** Integral of the total energy flux $\varepsilon_{\text{tot}}$ through the outer boundary of the computational region versus time $t$ for calculations [1–3].

It may also be noted that an explosive synthesis of radioactive nickel $^{56}_{28}\text{Ni}$ is possible behind the shock front. Using the simple reasoning behind explosive nucleosynthesis (Thielemann *et al.* 1990), we may assert that the $^{56}_{28}\text{Ni}$ nuclide is predominantly produced among the iron-peak elements if the local temperature exceeds its critical value $T_{\text{cr}}$. In the presupernova shells composed of $\alpha$-particle nuclei, radioactive nickel is synthesized outside the iron core (the shells of $^{28}_{14}\text{Si}$, $^{16}_{8}\text{O}$, $^{12}_{6}\text{C}$, etc.) in very short hydrodynamic times when the nuclear statistical equilibrium conditions are established. However, as was shown above, the post-shock temperature drops below its critical value by a time of $t \approx 0.1$ s. The radius of the shock front is $r_{\text{sw}} \approx 2.5$ (see above); i.e., it is smaller than the initial radius of the iron core $R_{\text{Fe}} = 4.38$. Nevertheless, we have reason to believe that the $^{56}_{28}\text{Ni}$ nuclide is still synthesized.

Indeed, having studied the entropy distribution for a toroidal atmosphere (Imshennik *et al.* 2003), we showed that the atmosphere for a given rotation law is predominantly formed from the matter of the silicon presupernova shell rather than from the outer iron core. Thus, if the toroidal atmosphere is assumed to have a silicon composition, then we can obtain a threshold estimate for the synthesized nickel mass. Estimating this mass is complicated by the difficulty of accurately determining the location of the contact boundary between the explosive destruction products of the low-mass neutron star and the matter of the toroidal atmosphere when using the Eulerian difference scheme. The location of the contact boundary on the equator can be roughly determined from the local minimum in the density and temperature distributions. Figure 5 shows the profiles of these thermodynamic quantities at $\theta = \pi/2$ as a function of the cylindrical radius $\tilde{r}$ for two early times, $t = 0.05$ s and $t = 0.09$ s. The sought location of the contact boundary on the equator can be determined from them: $r_{\text{cb}}(0.05) \approx 1.35$ and $r_{\text{cb}}(0.09) \approx 1.70$. At the early times under consideration, the shock front is nearly spherical in shape. In addition, the toroidal atmosphere is located near the equatorial plane, i.e., at angles $\theta > \pi/4$, and, hence, the behavior of the contact boundary far from the equator affects only slightly the synthesized nickel mass being estimated. Therefore, it would be natural to assume that outside the equator, the section of the contact surface by the plane passing through the rotation axis is a circumference with the center at $\tilde{r}(t) = r_{\min}$ and the radius $r(t) = r_{\text{cb}}(t) - r_{\min}$. The synthesized nickel mass of interest is calculated in the region of space outside this toroidal volume. Table 2 gives the masses of the atmospheric matter whose local temperature (behind the shock front) exceeds $T_{\text{cr}}$ for several early times. Clearly, the maximum mass in Table. 2 is an upper limit on the synthesized $^{56}_{28}$Ni mass. This result is in satisfactory agreement with the estimates by Zabrodina and Imshennik (2000): $M_{Ni} = 0.0185 M_\odot$ for the iron core radius $R_{Fe} = 4.38$ and $M'_{Ni} = 0.0277 M_\odot$ for $R_{Fe} = 2.37$.

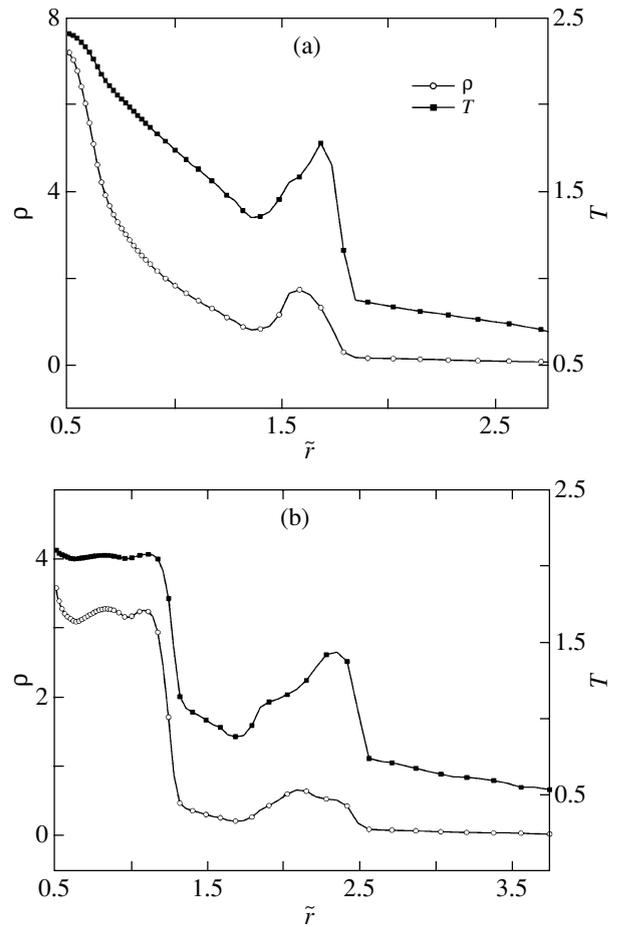

**Fig. 5.** Profiles of the density $\rho$ and temperature $T$ as a function of the cylindrical radius $\tilde{r}$ at $\theta = \pi/2$ for calculation [1] and for the times (a) $t = 0.05$ s and (b) $t = 0.09$ s.

**Table 2.** Masses of the atmospheric matter whose local temperature (behind the shock front) exceeds $T_{cr}$ at various times

| $t$, c | $M_{T>T_{cr}}$, $M_\odot$ |
|---|---|
| 0.03 | 0.011 |
| 0.04 | 0.021 |
| 0.05 | 0.017 |
| 0.07 | 0.008 |
| 0.09 | 0.000 |

## CONCLUSIONS

In the hydrodynamic theory of asymmetric supernova explosions, this work serves as a supplement that can be developed in terms of the rotational explosion scenario for collapsing supernovae (Aksenov *et al.* 1997; Imshennik and Zabrodina 1999; Zabrodina and Imshennik 2000). In this case, the important assumption of an axisymmetric explosion remains valid, because so far we have had to restrict our analysis to two-dimensional hydrodynamic models. The extreme complexity of the passage to three-dimensional models, which, of course, are the only ones that completely fit the rotational explosion scenario for collapsing supernovae, prompted us to investigate another hydrodynamic explosion model while remaining within the framework of the axisymmetric two-dimensional problem. In essence, this is because certain progress in describing the structure of the distribution of matter around a collapsed iron core has been made previously (Imshennik and Manukovskii 2000; in particular, Imshennik *et al.* 2003). The presence of stationary toroidal iron atmospheres instead of the rough assumption made in previous papers about a uniform iron gas distribution, which is essentially nonstationary on the long evolution time scales of neutron-star binaries considered in the above scenario, made this additional series of hydrodynamic calculations appropriate.

At the same time, much in the formulation of the problem remains unchanged. First, the unique choice of initial parameters of the circular orbit for an exploding low-mass neutron star with a critical mass is preserved thanks to the assumption that the kick velocity of the high-mass component in the binary (a pulsar) is 1000 km s$^{-1}$, in agreement with the observed high velocities of young pulsars. Second, the same equation of state in the approximation of nuclear statistical equilibrium is used for the explosion products of the neutron star and the surrounding gas of the toroidal atmosphere; i.e., in particular, the possible large expenditure of energy on the dissociation of iron nuclei into free nucleons is taken into account under the justified conservation condition for the neutron-to-proton number ratio, 30/26, typical of the $^{56}$Fe nuclides (Imshennik and Zabrodina 1999). Nevertheless, the two-dimensional peculiarities of the axisymmetric model under consideration make it necessary to change the direction of the $z$ coordinate axis to a perpendicular direction that coincides with the rotation axis of the toroidal atmosphere obtained; in turn, the latter is clearly the given rotation axis of the entire star before its collapse. Instead of describing an exploding neutron star in the shape of a sphere, we had to specify it in the shape of a torus in this case. It is qualitatively clear that this change in initial conditions excluded the possibility of the development of a directed asymmetry with the leading direction of the velocity vector of the exploded neutron star in the problem. In this model, the explosion is attributable only to energy release as the low-mass neutron star is destroyed. In this case, there is absolutely no contribution from the kinetic energy of the translational motion of the exploding star, because the orbital velocity becomes the rotational velocity of the torus introduced in the initial conditions for which the corresponding centrifugal force is exactly balanced by the attractive force of the pulsar placed at the coordinate origin. In this formulation of the problem, we rigorously took into account the gravitational interaction that was not included in previous papers on the hydrodynamic theory of asymmetric explosions at all. Thus, it is quite clear that the final explosion energy as the total energy that passed through the outer boundary of the computational region of the divergent shock wave will be appreciably lower than the energy obtained previously by Aksenov *et al.* (1997) and Imshennik and Zabrodina (1999), which is approximately equal to a characteristic value of $10^{51}$ erg. The main calculation of this paper, which is close in its final energy release to $10^{51}$ erg (the second column of Table 1), should be used for comparison. Indeed, the final energy is $\sim 0.2 \times 10^{51}$ erg (see Fig. 4). The leading sector of the shock wave in this calculation is located in the axial direction, which is attributable to a decrease in the matter density there rather than to the directed motion of the exploded neutron star, as was the case in previous models. Nevertheless, the numerical solution of this problem undoubtedly yielded a considerable lower limit for the final explosion energy. This is of great importance, because, first, it is still far from the characteristic supernova explosion energy ($\sim 10^{51}$ erg) obtained rigorously using the hydrodynamic theory for SN 1987A (Blinnikov 1999; Utrobin 2004) and, second, it is much higher than its value in the one-dimensional spherically symmetric hydrodynamic models of collapsing supernovae. For example, in the model by Imshennik and Nadyozhin (1977), this energy was found to be only $3 \times 10^{46}$ erg due to the rotation effect that was taken into account in the form of a centrifugal force averaged over the polar angle.

In discussing our results, we repeatedly touched on the closeness of the physical parameters obtained here to the parameters of the so-called model of a weak explosion (with half the initial internal energy) from Zabrodina and Imshennik (2000). The final explosion energy in the cited paper is even slightly larger than $10^{51}$ erg (see Fig. 6 from this paper). However, the following critical remark regarding this parameter can be made. The initial energy of the electron component of the degenerate iron gas, $\sim 0.4 \times 10^{51}$ erg, contributes significantly to it. This value just corresponds to a large mass of this gas, $\sim M_\odot$ (with the density $\rho = 5.66 \times 10^5$ g cm$^{-3}$ and the mean shock front radius $R_{\rm sw} \simeq 10^9$ cm), while this mass is an order of magnitude lower, $\sim (0.1-0.2) M_\odot$, under the formation conditions of a toroidal atmosphere. Accordingly, the energy contribution from this effect should be disregarded, so a value of $\sim 0.5 \times 10^{51}$ erg was actually obtained in the model of a weak explosion. The results under discussion must be compared with it; otherwise, they indeed differ little.

Our comparison suggests that the hydrodynamic model under consideration demonstrates an appreciable attenuation of the divergent shock wave, although, strictly speaking, the direct (see above) comparison of our results in the equatorial plane with those of the previous hydrodynamic model in the axial direction is conditional. Only a three-dimensional model will probably allow us to establish which of the two-dimensional hydrodynamic models being compared is suitable. Of course, in such a three-dimensional model, it will be unnecessary to "spread" the exploding neutron star into a torus, because the natural spherical shape adopted by Aksenov *et al.* (1997) and Imshennik and Zabrodina (1999) may be preserved in this case.

## ACKNOWLEDGMENTS

We thank M.S. Popov for assistance in tabulating the equation of state and for a helpful discussion of our results. This work was supported in part by the Russian Foundation for Basic Research (project no. 00-15-96572) and the Federal "Research and Development on Priority Fields of Science and Technology" Program (contract no. 40.022.1.1.1103) and an additional agreement no. 1 from January 31, 2003.

*Translated by V. Astakhov*